\documentclass[superscriptaddress,groupedaddress,nofootnoteinbib,12pt]{article}
\usepackage{color}
\usepackage{graphicx}
\usepackage{dcolumn}
\usepackage{bm}
\usepackage{amssymb}
\usepackage{amsmath}
\usepackage{sectsty}
\usepackage{colortbl}
\usepackage{latexsym}
\usepackage{float}
\usepackage{ifthen}
\usepackage{caption,subfig}
\usepackage{enumerate}
\usepackage{url}
\usepackage{caption,subfig}	
\usepackage{jcappub}

\newcommand{\eff}{{\rm eff}}

\def\be{\begin{equation}}
\def\ee{\end{equation}}
\def\ba{\begin{eqnarray}}
\def\ea{\end{eqnarray}}
\def\beq{\begin{eqnarray}}
\def\eeq{\end{eqnarray}}

\def\mpl{M_{\rm Pl}}

\def\d{\mathrm{d}}

\def\K{{\cal K}}

\def\L*{{\cal L}_*}
\def\L{\mathcal{L}}
\def\({\left(}
\def\){\right)}

\def\<{\langle}
\def\>{\rangle}

\def\cs2{c_{s}^{2}}

\def\be{\begin{equation}}
\def\ee{\end{equation}}
\def\ba{\begin{eqnarray}}
\def\ea{\end{eqnarray}}
\def\beq{\begin{eqnarray}}
\def\eeq{\end{eqnarray}}

\def\mpl{M_{\rm Pl}}

\def\d{\mathrm{d}}

\def\K{{\cal K}}

\def\L*{{\cal L}_*}
\def\L{\mathcal{L}}
\def\({\left(}
\def\){\right)}

\def\<{\langle}
\def\>{\rangle}

 \def\be   {\begin{equation}}   \def\ee   {\end{equation}}

 \def\ba  {\begin{eqnarray}}   \def\ea  {\end{eqnarray}}

\renewcommand{\thesubsection}{\arabic{section}.\arabic{subsection}}

\begin{document}
\hspace{5.2in} \mbox{IPMU14-0306}\\\vspace{-1.03cm} 

\title{Cosmological perturbations in massive gravity with doubly coupled matter}

\author{A. Emir G\"umr\"uk\c c\"uo\u glu$^{a}$, Lavinia Heisenberg$^{b,c}$, Shinji Mukohyama$^{d,e}$}
\affiliation{$^{a}$School of Mathematical Sciences, University of Nottingham, \\
University Park, Nottingham, NG7 2RD, UK}
\affiliation{$^{b}$Nordita, KTH Royal Institute of Technology and Stockholm University, \\
Roslagstullsbacken 23, 10691 Stockholm, Sweden}
\affiliation{$^{c}$Department of Physics \& The Oskar Klein Centre, \\
AlbaNova University Centre, 10691 Stockholm, Sweden}
\affiliation{$^{d}$Kavli Institute for the Physics and Mathematics of the Universe,\\
Todai Institutes for Advanced Study, University of Tokyo (WPI),\\
5-1-5 Kashiwanoha, Kashiwa, Chiba 277-8583, Japan}
\affiliation{$^{e}$Yukawa Institute for Theoretical Physics, Kyoto University, \\
Kyoto 606-8502, Japan}

\emailAdd{Emir.Gumrukcuoglu@nottingham.ac.uk} 
\emailAdd{Lavinia.Heisenberg@unige.ch}
\emailAdd{shinji.mukohyama@ipmu.jp}

\abstract{We investigate the cosmological perturbations around 
 FLRW solutions to non- linear massive gravity with a new
effective coupling to matter proposed recently. 
Unlike the case with minimal matter coupling, all five
degrees of freedom in the gravity sector propagate on generic
self-accelerating FLRW backgrounds. 
We study the stability of the cosmological solutions and put constraints
on the parameters of the theory by demanding the correct sign for the
kinetic terms for scalar, vector and tensor perturbations.}

\maketitle

\section{Introduction}

Albert Einstein proposed his breakthrough theory of General Relativity (GR) in 1916. Since then it has successfully resisted the largest variety of tests. GR triumphs in shaping our current understanding of physics from millimeter length scales all the way up to cosmological scales. Besides its remarkable successes, the theory suffers also from a number of confrontations from both theoretical and observational perspectives on very large and small scales. The unification of gravity with quantum mechanics beyond the validity of the effective field theory description remains a fundamental open question over decades. Similarly, requiring the unification of gravity with the standard model of particle physics indicates beyond doubt the need for a modification of the theory in the UV. Furthermore, the unavoidability of cosmic and black hole singularities signals that the theory in its original form might be inadequate for gravitational phenomena at very high energies. On the other hand, the observational confrontations at the other end of the energy spectrum, like Cosmological Constant (CC) problem \cite{Weinberg:1988cp} and the recent accelerated expansion of the Universe  \cite{Perlmutter:1997zf,Riess:1998cb,Tonry:2003zg} might be seen as an indication of the breakdown of gravity on these large cosmological scales.\\

The aforementioned problems have initiated the study on modifications of gravity in the IR and UV. The renormalization properties of GR can be improved by abandoning Lorentz invariance at high energies as in Ho\v{r}ava-Lifshitz gravity \cite{Horava:2009uw,Blas:2009yd,Blas:2009qj,Mukohyama:2009zs,Blas:2010hb,Nojiri:2010wj,Mukohyama:2010xz} (see also references therein). The existence of a preferred time foliation renders the gravitational theory power-counting renormalizable due to addition of higher spatial derivatives without adding higher-order time derivatives. Ho\v{r}ava-Lifshitz gravity could be an appropriate framework of constructing a sensible theory of quantum gravity once the renormalizability has really been proven and the relativistic low energy limit is successfully implemented. Similarly, the hope to find a regularization scheme for the singularities in gravity led to considerations of Born-Infeld inspired gravity models, which represent a specific infinite order higher curvature modification of GR  \cite{Deser:1998rj,Ketov:2001dq,Wohlfarth:2003ss,Pani:2012qb,Scargill:2012kg,Olmo:2013gqa,Odintsov:2014yaa,Komada:2014asa,Jimenez:2014fla}. 
However, this type of modifications with higher order curvature terms usually introduces additional degrees of freedom, which in turn lead to a loss of unitarity despite being renormalizable. A possible rescue might come from  formulating the theory {\`a} la Palatini, which opens up some avenues to avoid the cosmological and black hole singularities \cite{Vollick:2005gc,Pani:2012qd,Jimenez:2014fla, Komada:2014asa}.\\

A promising infra-red modification of GR is the non-linear extension of massive gravity proposed by de Rham-Gabadadze-Tolley (dRGT) \cite{deRham:2010ik,deRham:2010kj}, which successfully extends the unique mass term at the linear level, the Fierz--Pauli action \cite{Fierz:1939ix,VanNieuwenhuizen:1973fi}, and is given by a very specific structure of a 2-parameter family interaction potential (for extensive reviews, see \cite{Hinterbichler:2011tt,deRham:2014zqa}). The form of the potential is designed such that the ghost--like sixth polarization of graviton, which leads to the Boulware--Deser (BD) instability \cite{Boulware:1973my}, is completely removed. The absence of the BD ghost has been proven in a multitude of languages and formalisms \cite{Hassan:2011hr,deRham:2011rn,Hassan:2011ea,deRham:2011qq,Mirbabayi:2011aa,Golovnev:2011aa,Hassan:2012qv,Hinterbichler:2012cn,Kluson:2012wf,Deffayet:2012zc}. The decoupling limit of the theory contains Galileon interactions \cite{Nicolis:2008in, deRham:2010ik, deRham:2010tw,Ondo:2013wka} and shares the property of non-renormalization theorem \cite{deRham:2012ew,Brouzakis:2013lla,deRham:2014wfa,Heisenberg:2014raa}. The quantum behavior of the theory has been further explored in \cite{Park:2010rp,Buchbinder:2012wb,deRham:2013qqa, deRham:2014naa}. There has been a flurry of investigations concerning the phenomenological aspects of the dRGT massive gravity, specially its cosmological implications \cite{deRham:2010tw,PhysRevD.84.124046,PhysRevLett.109.171101,deRham:2011by,Chamseddine:2011bu,Koyama:2011xz,Koyama:2011wx,Gumrukcuoglu:2011zh,Gratia:2012wt,Vakili:2012tm,Kobayashi:2012fz,Fasiello:2012rw,Volkov:2012zb,Tasinato:2012ze,DeFelice:2013bxa,Fasiello:2013woa,Heisenberg:2014kea,Comelli:2013tja,Motloch:2014nwa}. From a technical stand point, the very specific potential of this non-linear massive gravity theory was constructed by demanding that the vanishing helicity-2 part ($h_{\mu\nu}=0$ ) of the interactions consists only of total derivatives for the helicity-0 field $\pi$. Hassan and Rosen realized that this criteria was automatically fulfilled by writing the interactions in terms of a deformed determinant which, expressed in terms of the antisymmetric Levi-Civita tensor, guarantees the total derivative nature of the interactions for the helicity-0 field \cite{Hassan:2011vm}. Very soon they proposed the bimetric extension of massive gravity, where the reference metric $f_{\mu\nu}$ becomes dynamical through the inclusion of the corresponding kinetic term \cite{Hassan:2011zd}. \\

In massive (bi-)gravity the existence of two metrics (in bigravity both metrics are dynamical whereas in massive gravity the second metric is a non-dynamical reference metric) comes hand in hand with the natural question of how to couple these two metrics in a consistent way to the matter sector without causing the BD ghost to reappear \cite{Khosravi:2011zi, Akrami:2012vf,Akrami:2013ffa,Tamanini:2013xia,Akrami:2014lja, deRham:2014naa,Yamashita:2014fga,Noller:2014sta,Hassan:2014gta,Schmidt-May:2014xla,Enander:2014xga}. Once this has been successfully constructed at the classical level, the same criteria needs to be maintained at the quantum level. In the case where the matter sector couples minimally to only one metric, the absence of BD ghost had been proven non-linearly at the classical level already in \cite{Hassan:2011zd}. Furthermore, in \cite{deRham:2014naa} it has been shown that this property remains true at the quantum level since the quantum corrections do not destabilize the form of the potential but rather gives rise to a cosmological constant. In bigravity there is nothing special about the one metric versus the other one, therefore it might be tempting to consider the case in which the matter field couples minimally to both metrics simultaneously. However, as shown in  \cite{deRham:2014naa} not only the BD ghost degree of freedom is present already at the classical level, the quantum corrections detune the potential structure at an arbitrarily low scale. Moreover, requiring that the specific ghost-free potential structure is maintained at the quantum level, a new effective metric can be constructed through which the matter field can couple to both metrics at the same time  \cite{deRham:2014naa}. This composite effective metric is very special: 
once the matter loops are integrated, the quantum corrections yield a contribution in the form of the allowed potential in the dRGT action. Another complementary analysis yields the same conclusion in \cite{Noller:2014sta}. In \cite{deRham:2014naa} it was shown that this specific coupling of the matter field to the effective metric does not excite any ghost degree of freedom at the strong coupling scale and therefore can be considered as an effective field theory with a cut-off above the strong coupling scale. \\

The present paper is devoted to the study of the cosmological perturbations in the massive gravity theory with a special coupling to the matter sector through a new composite effective metric. The FLRW dynamics were already studied in \cite{deRham:2014naa} where this interaction was proposed for the first time. We aim to push forward these preliminary studies and investigate in detail the behavior and stability of perturbations. After briefly reviewing the dRGT massive gravity together with the new effective composite metric that the matter sector couples to in Section \ref{sec:dRGT}, we explore the background evolution on a homogeneous and isotropic space-time in Section \ref{sec:background_evolution}. The new coupling establishes a framework in which the no-go theorem for FLRW solutions can be avoided. Section \ref{sec:perturbations} is consecrated to a detail study of the stability of the perturbations. Finally, we summarize our results in section \ref{sec:conclusion}.\\

Throughout the paper, we will adapt to the units in which $c=\hbar=1$ and use the notation $M_{\rm Pl}=1/\sqrt{8 \pi G}$ for the reduced Planck mass. Furthermore, we follow the mostly plus metric signature convention. Some short-cut notations are used to denote the contractions of rank-2 tensors ${\cal K}^{\mu}_{~\mu}=[{\cal K}]$,~ ${\cal K}^{\mu}_{~\nu}{\cal K}^{\nu}_{~\mu}=[{\cal K}^2]=({\cal K}_{\mu\nu})^2$,~ ${\cal K}^{\mu}_{~\nu}{\cal K}^{\nu}_{~\rho}{\cal K}^{\rho}_{~\mu}=[{\cal K}^3]=({\cal K}_{\mu\nu})^3$~etc. Greek indices run form $0$ to $3$ while Latin indices from $1$ to $3$. With the latin indices we denote contractions in the same way as for the Greek indices $h_{ij}h^{ij}=(h_{ij})^2$, $A_i A^i=(A_i)^2$...etc.


\section{dRGT massive gravity with doubly coupled matter}
\label{sec:dRGT}
In this section we will first review the ghost-free interactions in the theory of massive gravity and setup the framework in which we will perform our analysis of cosmological perturbations. Our starting point is the action for massive gravity and the matter action where the ordinary matter fields still couple minimally to the physical metric $g$ and an additional scalar field $\chi$ couples to the composite effective metric proposed in \cite{deRham:2014naa}

\begin{equation}\label{action_MG_effcoupl}
\mathcal{S} = \int \mathrm{d}^4x \big[ \frac{\mpl^2}{2} \sqrt{-g}\left(R[g]-\frac{m^2}{2}\sum_n \alpha_n{\cal U}[\cal K]  +\mathcal{L}_{\rm matter} \right)+\sqrt{-g_{\rm eff}}\mathcal{L}_\chi(g_{\rm eff},\chi)\big]\,,
\end{equation}
where the potential interactions are given by \cite{deRham:2010ik,deRham:2010kj}
\begin{eqnarray}
\mathcal{U}_2[\mathcal{K}] &=& \mathcal{E}^{\mu\nu\rho\sigma}  \mathcal{E}^{\alpha\beta}_{\;\;\;\;\; \rho\sigma} \K_{\mu\alpha} \K_{\nu\beta} = 2\left( [\K]^2-[\K^2]\right), \nonumber\\
\mathcal{U}_3[\mathcal{K}] &=& \mathcal{E}^{\mu\nu\rho\sigma}  \mathcal{E}^{\alpha\beta\kappa}_{\;\;\;\;\;\;\; \sigma} \K_{\mu\alpha} \K_{\nu\beta}  \K_{\rho\kappa}=[\K]^3-3[\K][\K^2]+2[\K^3],  \nonumber\\
\mathcal{U}_3[\mathcal{K}] &=& \mathcal{E}^{\mu\nu\rho\sigma}  \mathcal{E}^{\alpha\beta\kappa\gamma} \K_{\mu\alpha} \K_{\nu\beta}  \K_{\rho\kappa}  \K_{\sigma\gamma} =[\K]^4-6[\K]^2[\K^2]+3[\K^2]^2+8[\K][\K^3]-6[\K^4],
\end{eqnarray}
where $\mathcal{E}$ stands for the Levi-Cevita tensor. The tensor $\K$ has a very non-trivial structure in form of a square root
\begin{equation}
\K^\mu _\nu[g,f] =\delta^\mu_\nu - \left(\sqrt{g^{-1}f}\right)^\mu_\nu \,,
\end{equation}
The mass potentials in \ref{action_MG_effcoupl} breaks the diffeomorphism invariance completely. However, we can restore it with the help of four St\"uckelberg fields $\phi^a$ by promoting the Minkowski reference metric to the space-time tensor
\begin{equation}\label{Stueckelbergfields}
f_{\mu\nu} \to \eta_{ab}\partial_\mu \phi^a \partial_\nu\phi^b\,.
\end{equation}
In \cite{deRham:2014naa} a new effective composite coupling with a particular combination of the metrics was proposed based on the question of which specific coupling would ensure that one-loop corrections from virtual matter fields do not detune the special structure of the potential. 
The matter Lagrangian $\mathcal{L}_\chi$ thus needs to couple to the effective metric which was found to have the following form
\begin{equation}
g^\eff_{\mu\nu} \equiv \alpha^2 g_{\mu\nu}+2\,\alpha\,\beta\, g_{\alpha\mu} \left(\sqrt{g^{-1}f}\right)^\alpha_{\nu} + \beta^2 f_{\mu\nu}\,.
\label{eq:geff}
\end{equation}
With this choice, the BD degree of freedom is not generated thanks to the form of the volume element for the effective metric 
\begin{equation}
\label{eq:detgeff}
\sqrt{-g_{\eff}}= \sqrt{-g}\  \det\(\alpha +\beta\left(\sqrt{g^{-1}f}\right)^\mu_\nu\)\,,
\end{equation}
which can be rewritten using the expansion of a deformed determinant \cite{Hassan:2011vm} as
\begin{equation}
\label{eq:detgeff2}
\sqrt{-g_{\eff}}= \sqrt{-g}\  \sum_{n=0}^4 \frac{(-\beta)^n}{n!}(\alpha+\beta)^{4-n} \mathcal{U}_n[K]\,.
\end{equation}
In other words, the volume element itself can be written in terms of the allowed potentials of dRGT theory. For sake of simplicity we will consider a scalar field $\chi$ with an arbitrary potential that couples to this effective metric via
\begin{equation}
\mathcal{L}_{\chi} =- \frac{1}{2}g_\eff^{\mu\nu}\,\partial_\mu\chi\partial_\nu\chi - V(\chi)\,.
\label{eq:chiaction}
\end{equation}
In the following, we neglect the tadpole contribution $\mathcal{U}_1$ to the mass term by setting $\alpha_1=0$, while the $\mathcal{U}_0$ term is absorbed into a bare cosmological constant, which plays the role of a place-holder for ordinary matter fields 
\begin{equation}
\mathcal{L}_{\rm matter}=-\mpl^2\Lambda\,.
\end{equation}


\section{Background evolution}\label{sec:background_evolution}

The original massive gravity theory without the effective coupling was subject to a no-go theorem for flat FLRW solutions \cite{PhysRevD.84.124046}, 
arising from the fact that the equation of motion for the St\"uckelberg field enforces the scale factor to be constant.
However, the presence of this effective metric through the matter coupling yields a significant modification of the equation of motion for the St\"uckelberg field and hence allows exact FLRW solutions with flat reference metric as pointed out in  \cite{deRham:2014naa}. We would like to push this analysis further and study how perturbations behave on top of this FLRW background. Our starting Ansatz for the dynamical metric is the homogeneous and isotropic flat FLRW 
\begin{equation}
ds_g^2=-N^2 dt^2 +a^2 \delta_{ij} dx^idx^j\,,
\end{equation}
while the non-dynamical metric is the pull-back of the Minkowski metric in the St\"uckelberg field space to the physical space-time, which we parametrize as
\begin{equation}
ds_f^2= f_{\mu\nu}dx^\mu dx^\nu = -\dot{f}^2 + a_0^2 \delta_{ij} dx^idx^j\,.
\label{eq:reference}
\end{equation}
This corresponds to the choice $\phi^0=f(t)$, $\phi^i = a_0 x^i$ for the St\"uckelberg fields introduced in (\ref{Stueckelbergfields}). For the case $f(t)=t$ and $a_0=1$, this coincides with the unitary gauge $\phi^a=x^a$. The advantages of this choice are i) the function $f(t)$ reintroduces the time reparametrization invariance and allows us to directly find the constraint implied by the St\"uckelberg fields, and as we show shortly, it also makes the integration of the equation of motion trivial; ii) the introduction of the second lapse function $n\equiv \dot{f}$ allows us to identify any explicit dependence on $N$ which cannot be removed by changing the time coordinate; iii) the introduction of $a_0$ makes the determination of the physical quantities which are independent of the scaling of the coordinates manifest. In this background, the effective metric defined in equation (\ref{eq:geff}) corresponds to the line element
\begin{equation}
ds^2_\eff = -N^2_\eff dt^2+a_\eff^2 \delta_{ij}dx^idx^j\,,
\end{equation}
where $N_{\eff}$ and $a_{\eff}$ are the effective lapse and scale factor respectively
\begin{equation}
N_\eff \equiv \alpha\,N+\beta\,\dot{f}\,,\qquad
a_\eff \equiv \alpha\,a+\beta\,a_0\,.
\end{equation}
Compatible with our above homogeneous and isotropic Ansatz we assume a $\chi$ field condensate depending only on time $\chi=\chi(t)$. For our convenience we introduce the following quantities
\begin{eqnarray}\label{shortcuts}
H&\equiv& \frac{\dot{a}}{a\,N}\,, \\
r&\equiv& \frac{\dot{f}/a_0}{N/a}\,, \\
U(A) & \equiv &6\,\sum_{n=2}^4\,\alpha_n (1-A)^n\,.
\end{eqnarray}
where the ratio of the scale factors is denoted by $A\equiv a_0/a$ and  $H$ is the expansion rate of the physical $g$ metric, while $r$ defines the light cone of the $f$ metric (or the speed of light propagating in the $f$ metric). Our action (\ref{action_MG_effcoupl}) in the mini-superspace becomes (up to total derivatives):
\begin{eqnarray}
\frac{S}{V} &=& \mpl^2\int dt \,a^3 N\,\Bigg\{-\Lambda-3H^2-m^2\left[U(A)+\frac{U_{,A}}{4}\left(r-1\right)A\right]\Bigg\} \nonumber\\
&&+\int dt \,a_\eff^3 N_\eff \left[ \frac{\dot{\chi}^2}{2\,N_\eff^2}-V(\chi)\right]\,,
\label{eq:minisuperspace}
\end{eqnarray}
where $U_{,A}=\partial_AU$. 
We are now ready to compute the background equations of motion by varying the action (\ref{eq:minisuperspace}) with respect to $N$, $a$, $\chi$ and $f$. We remark that the resulting system of equations of motion contains a redundant equation, as they are connected through the contracted Bianchi identity,
\begin{equation}
\frac{\partial}{\partial t} \frac{\delta S}{\delta N} - \frac{\dot{a}}{N}\frac{\delta S}{\delta a}- \frac{\dot{\chi}}{N}\frac{\delta S}{\delta \chi}- \frac{\dot{f}}{N}\frac{\delta S}{\delta f} =0\,.
\label{eq:bianchi}
\end{equation}
In the remainder of the paper, we use the following functions instead of the parameters $\alpha_n$ 
\begin{equation}
\rho_m (A)\equiv U(A)-\frac{A}{4}\, U_{,A} \,,\qquad
J(A)\equiv \frac{1}{3}\,\partial_A \rho_m(A)\,,\qquad
Q(A) \equiv \frac{1}{4}U_{,A}\,.
\end{equation}
It will turn out that $\rho_m$ denotes the dimensionless effective energy density from the mass term. First of all, the Friedmann equation can be calculated by varying the action (\ref{eq:minisuperspace}) with respect to the lapse $N$, which yields
\begin{equation}
3\,H^2 = \Lambda + m^2 \rho_m +\frac{\alpha\,a_\eff^3}{\mpl^2\,a^3}\left(\frac{\dot{\chi}^2}{2\,N_\eff^2}+V\right)\,,
\label{eq:eqN}
\end{equation}
Similarly, the variation of the mini-superspace action (\ref{eq:minisuperspace}) with respect to the scale factor $a$ and combining it with the Friedmann equation, gives rise to the acceleration equation 
\begin{equation}
\frac{2\,\dot{H}}{N}= m^2\,J\,A\,(r-1) - \frac{\alpha\,a_\eff^3}{\mpl^2a^3}\left[
\left(1+\frac{N_\eff/a_\eff}{N/a}\right)\frac{\dot{\chi}^2}{2\,N_\eff^2}+ \left(1- \frac{N_\eff/a_\eff}{N/a}\right) V\right]\,.
\label{eq:eqa}
\end{equation}
Before moving on, a few comments are in order. From (\ref{eq:geff}), we see that for $\alpha=0$, the $\chi$ field couples only to the $f$ metric. Indeed, in this case, equations (\ref{eq:eqN}) and (\ref{eq:eqa}) correspond to their counterparts in dRGT theory. Conversely, in the $\beta=0$ case, the $\chi$ field couples to $\alpha g_{\mu\nu}$. Thus the terms containing the $\chi$ field in (\ref{eq:eqN}) and (\ref{eq:eqa}) correspond to a canonical scalar field on massive gravity with coupling constant $\mpl^2/\alpha^4$.\\
The equation of motion for the $\chi$ field is just the standard conservation equation for a field minimally coupled to the $g_\eff$ metric, namely,
\begin{equation}
\frac{1}{N_\eff}\,\frac{\partial}{\partial t}\,\left(\frac{\dot{\chi}}{N_\eff}\right) +3\frac{\dot{a}_\eff}{a_\eff\,N_\eff}\,\frac{\dot{\chi}}{N_\eff}+V'(\chi)=0\,.
\label{eq:eqchi}
\end{equation}
Last but not least, the variation with respect to the St\"uckelberg field results in
\begin{equation}
m^2\,\mpl^2J=\frac{\alpha\beta\,a_\eff^2}{a^2}\left(\frac{\dot{\chi}^2}{2\,N_\eff^2}-V\right)\,.
\label{eq:eqf}
\end{equation}
The above equation is the key property of this model. It provides a constraint on the background evolution, and can be interpreted as an algebraic relation to determine $N$, $N_\eff$, $r$, $\dot{f}$.  
Note also, that it is a quadratic equation with two solutions. On the other hand, since the same combination $\dot{\chi}^2/N_\eff^2$ appears in the Friedmann equation, the above constraint can be used without ambiguity on (\ref{eq:eqN}) to obtain:
\begin{equation}
3\,H^2=\Lambda +m^2\left[\rho_m+J\left(A+\frac\alpha\beta\right)\right]+\frac{2\,\alpha\,a_\eff^3}{a^3}\,\frac{V}{\mpl^2}\,.
\end{equation}
Moreover, the mini-superspace action (\ref{eq:minisuperspace}) depends only on the first derivative of the temporal St\"uckelberg field $f$, hence we can trivially integrate equation (\ref{eq:eqf}) once to obtain:
\begin{equation}
m^2\mpl^2\,a^3\,Q+ \beta a_\eff^3\left(\frac{\dot{\chi}^2}{2\,N_\eff^2}+V\right)=a_0^3\,m^2 \mpl^2\kappa\,,
\label{eq:eqf-alt}
\end{equation}
where $\kappa$ is a dimensionless integration constant, independent of the normalization of the scale factor. By combining this with the Friedmann equation (\ref{eq:eqN}), we get
\begin{equation}
3\,H^2=\Lambda + m^2\left[\rho_m-\frac{\alpha}{\beta} \left(Q-\kappa\,A^3\right)\right]\,.
\label{eq:friedmann-final}
\end{equation}
We would like to emphasize that the contributions coming from the mass term and the $\chi$ field on the right hand side of the above equation, contain terms up to $A^{3}$, i.e. with the fastest redshift being $a^{-3}$. We also remark that we did not assume any approximation in order to obtain the above equation, however the resulting Friedmann equation does depend on the information of initial conditions, encoded in the integration constant $\kappa$. 
Furthermore, one can remove the explicit $\chi$ dependent part in the acceleration equation (\ref{eq:eqa}): upon using the equations (\ref{eq:eqf}) and (\ref{eq:eqf-alt}) to eliminate the kinetic and potential terms for the $\chi$ field, it reads
\begin{equation}
\frac{2\,\dot{H}}{N} = -m^2\,\left[J\,A - \frac{\alpha}{\beta}\left(Q-\kappa\,A^3-J\right)
\right]\,.
\label{eq:acceleration-final}
\end{equation}
Note that this equation is nothing else but the time derivative of equation (\ref{eq:friedmann-final}) (up to a factor of $3\,H\,N$). This can be easily checked once the following relations have been used
\begin{equation}
\dot{\rho}_m = -3\,N\,H\,J\,A \,,\qquad
\dot{Q} = 3\,N\,H\,(J-Q) \,,\qquad
\dot{A} = -N\,H\,A\,. 
\end{equation}
Any $\chi$ dependence in the further steps can be removed by using equations (\ref{eq:eqf}), (\ref{eq:eqf-alt}) and their derivatives. It is straightforward to verify that these equations and their derivatives are already consistent with the $\chi$ field equation of motion (\ref{eq:eqchi}), through the contracted Bianchi identity (\ref{eq:bianchi}).\\
We end this discussion with an example, namely the minimal model with parameters 
$\alpha_2=1$, $\alpha_3=\alpha_4=0$ and $m^2>0$. In this case, the Friedmann equation becomes
\begin{equation}
3H^2 = \Lambda + 3\,m^2 (A-1)\left(A-2 - \frac{\alpha}{\beta}\right) + \frac{m^2 \alpha\,\kappa\,A^3}{\,\beta}
\end{equation}
At late times, $A \ll 1$, the universe is dominated by the effective cosmological constant $\Lambda_\eff \equiv \Lambda +3 m^2 (\alpha+2\,\beta)/\beta$. Conversely, at early times, we have $A\gg 1$ and the integration constant term dominates the expansion, mimicking pressureless dust. Requiring positive effective energy density imposes $\alpha\kappa/\beta >0$.



\section{Stability of the perturbations}\label{sec:perturbations}
In the previous section we have analyzed the theory on a flat FLRW space-time and saw that the new coupling of the matter sector with the effective metric circumvents the no-go theorem for the existence of flat FLRW solutions in the massive gravity theory. Of course the original no-go theorem is not a failure for massive gravity. In fact one can construct cosmological solutions that are inhomogeneous and/or anisotropic on very large scales but mimic arbitrarily close FLRW solutions within distances smaller than horizon \cite{PhysRevD.84.124046}. However, if one insists on flat FLRW solutions on all scales, the new effective composite metric enables us to accommodate such solutions. In \cite{deRham:2014naa} it has been explicitly shown, that the Boulware-Deser ghost is absent around these exact FLRW solutions. An additional analysis in the decoupling limit has indicated that on scales below the strong coupling scale $\Lambda_3^3=\mpl m^2$ the theory avoids the Boulware-Deser ghost and hence can be considered as a perfectly valid effective field theory. Beyond the strong coupling scale the theory admits a Boulware-Deser ghost whose existence was proven perturbatively in \cite{deRham:2014naa} but also in a more general setup non-perturbatively in \cite{deRham:2014fha}. Nevertheless, the absence of the ghost at the very least till the strong coupling scale makes the theory still very attractive for phenomenological studies as long as one remains within the regime of validity of the effective field theory. The action with the effective matter coupling \ref{action_MG_effcoupl} consist of an infinite number of operators some of which enter at the scale $\Lambda_3^3$, some of which enter at the cutoff scale $\Lambda_{\rm cut-off}$ and some of which enter in between these
two scales. The operators below $\Lambda_{\rm cut-off}$ are just fine whereas the ones entering at and above $\Lambda_{\rm cut-off}$ are not to be trusted as they contain a ghost. For any solution whose physical scales (all the fields and their derivatives) are smaller than the cut-off, it is simply impossible that this solution would have excited the operators which are at or above the cut-off, meaning that these solutions would not rely on the ghost. Let us emphasize again, that the FLRW solutions found in  \cite{deRham:2014naa} do not excite the Boulware-Deser ghost, i.e. all the ghostly-like operators disappear, giving rise to a healthy theory on that background.

The central goal of this work is to determine the stability of perturbations around these solutions. For this purpose, let us consider the following perturbations for the dynamical metric $g_{\mu\nu}$
\begin{eqnarray}\label{perturbed_metric}
\delta g_{00} &=& -2\,N^2\,\Phi\,,\nonumber\\
\delta g_{0i} &=& N\,a\,\left(\partial_i B+B_i\right)\,,\nonumber\\
\delta g_{ij} &=& a^2 \left[2\,\delta_{ij}\psi +\left(\partial_i\partial_j-\frac{\delta_{ij}}{3}\partial^k\partial_k\right)E+\partial_{(i}E_{j)}+h_{ij}\right]\,.
\end{eqnarray}
where all perturbations are functions of time and space and accord to the transformations under spatial rotations. Note that $\delta^{ij}h_{ij} = \partial^ih_{ij} = \partial^i E_i = \partial^i B_i=0$. Furthermore, we choose to keep the St\"uckelberg fields purely background, thus fixing the gauge freedom completely. The perturbed reference metric is thus still given by (\ref{eq:reference}). 
We perturb the scalar field $\chi$ as follows
\begin{equation}
\chi=\chi_0(t)+\mpl \delta\chi\,.
\end{equation}
In this setup, the action in \ref{action_MG_effcoupl} contains na\"ively counted eleven
degrees of freedom (dof) two of which are traceless symmetric spatial tensor fields ($h_{ij}$), four of which are divergence-free spatial vector fields ($B_i$, $E_i$) and the remaining five dof are scalars ($\Phi$, $B$, $\psi$, $E$, $\delta\chi$) with no remaining gauge symmetries. However, out of these dof two of the scalar fields ($\Phi$, $B$) and two of the vector fields ($B_i$) are non-dynamical. Furthermore, since the Boulware-Deser ghost is absent in this configuration as shown in \cite{deRham:2014naa}, the dRGT tuning along with the specific coupling of the $\chi$ field will allow us to integrate out one more combination. At the end, we will be left with two tensor, two vector and two scalar dof. These correspond to the five polarizations of the massive spin--2 field and the matter field $\chi$ perturbations. In the following we will present the analysis of the perturbations for each sector independently. We will investigate the question of whether or not the perturbations contain ghost and/or Laplacian instability. A ghost is a field with negative kinetic energy, i.e with the wrong sign kinetic term. On the other hand Laplacian instability indicates the presence of negative squared propagation speed. Additionally, a tachyon represents an instability in the potential.

\subsection{Tensor perturbations}
Tensor perturbations are the transverse traceless part of the metric fluctuations and 
are the only sources of gravitational waves in GR.
When a plane gravitational wave perturbations passes through space-time, it stretches the space in a way that a circle in the plane is distorted into an ellipse. Even if their direct detection is still lacking an indirect evidence for their existence is provided by pulsar binary system \cite{Weisberg:1981mt}. A general modified gravity model comes hand in hand with two potential effects on the equation for these gravitational waves. First of all, the propagation speed of the gravitational wave can be different from the speed of light. In this respect, we will demand the absence of Laplacian instability. Of course the propagation speed of light can be sub- (super-) luminal even though observational constraints will pin down large discrepancies to the speed of light. Second of all the modification of gravity will alter the friction term in the equation for the tensor perturbations. Needless to say that all these modifications will give rise to potential differences in the observations and the parameters of the theory have to be constrained in a way that do not contradict the real observations. In order to study the stability of the tensor perturbations we will decompose the tensor field in Fourier modes with respect to the spatial coordinates since we are working on a homogeneous background metric with no spatial curvature
\begin{equation}\label{tensor_Fourier}
h_{ij}=\int \frac{\d^3k}{(2\pi)^{3/2}}h_{ij,\vec{k}}(t) \exp(i\vec{k}\cdot\vec{x}) +c.c\,.
\end{equation}
After plugging our Ansatz \ref{perturbed_metric} for the metric perturbations into our Lagrangian \ref{action_MG_effcoupl}, decomposing the tensor field as in \ref{tensor_Fourier} and using the background equations, the action quadratic in the tensor perturbations becomes (up to boundary terms)
\begin{equation}
S^{(2)}_{\rm tensor} = \frac{\mpl^2}{8}\int d^3k\,dt\,N\,a^3\,\left[\frac{1}{N^2}\dot{h}_{ij,\vec{k}}^\star \dot{h}^{ij}_{\vec{k}}-\left(\frac{k^2}{a^2}+m_{T}^2\right)h_{ij,\vec{k}}^\star h^{ij}_{\vec{k}} \right]\,,
\end{equation}
where the mass of the tensor perturbations is given by
\begin{equation}
m_{T}^2\equiv \frac{m^2\,A\,(r-1)}{A-1}\Bigg[JA^2\frac{a}{a_\eff}\left(\frac{\alpha}{A^2}+2\frac{(\alpha+\beta)}{A}+\beta\right)-\frac{Q+A^2\,\rho_m}{(A-1)^2}\Bigg]\,.
\label{eq:MGW2}
\end{equation}
The tensor modes have already the right sign for the kinetic term. Similarly, they do not exhibit gradient instabilities either. The only concern might come from the fact that for $m_{T}^2<0$, there will be a tachyonic instability. However, the time-scale of the instability is of the order of inverse graviton mass, meaning that for a graviton mass of order of $H$ today, it takes the age of the universe to develop such a tachyonic instability.

\subsection{Vector perturbations}
The discovery of the alignment of the low multipoles of the CMB and the hemispherical asymmetry could indicate the existence of a privileged
direction in the universe. This has motivated the exploration of cosmic vector fields. They naturally arise in modified gravity, like in massive gravity. 
In this subsection, we will study the stability conditions of the vector perturbations after integrating over the non-dynamical vector modes. 
In a similar way as for the tensor perturbations, we will decompose the vector modes $E_i$ and $B_i$ in their Fourier modes
\begin{eqnarray}
E_i=\int \frac{d^3k}{(2\pi)^{3/2}} E_{i,\vec{k}}\;(t)e^{i\vec{k}\cdot\vec{x}} +c.c, \;\;\;\;\;\;\;\;\;\;\;     B_i=\int \frac{d^3k}{(2\pi)^{3/2}} B_{i,\vec{k}}\;(t)e^{i\vec{k}\cdot\vec{x}} +c.c.
\label{fourierEandB}
\end{eqnarray}
We will then expand the action up to second order in the vector perturbations, which yields 
\begin{equation}\label{perturbed_vector}
S^{(2)}_{\rm vector} = \frac{\mpl^2}{16}\int d^3k\,N\,dt\,k^2a^3 \left[
\frac{1}{N^2}\dot{E}_{i,\vec{k}}^\star\dot{E}^i_{\vec{k}}-\frac{2}{a\,N}\left(\dot{E}_{i,\vec{k}}^\star B^i_{\vec{k}}+B_{i,\vec{k}}^\star \dot{E}^i_{\vec{k}}\right)-m_{T}^2 E_{i,\vec{k}}^\star E^i_{\vec{k}} 
+\frac{4}{a^2}m_V^2\,B_{i,\vec{k}}^\star B^i_{\vec{k}}
\right]\,,
\end{equation}
where we defined for convenience the shortcut notation
\begin{equation}
m_V^2 \equiv 
1+\frac{4\,\beta\,a^4A^2}{k^2\,a_\eff^2(1+r)A}\left(\frac{N_\eff}{N(r+1)}+\frac{a_\eff}{a}\right)\left(-\frac{\dot{H}}{N}\right)\,.
\label{eq:veckin}
\end{equation}
As we mentioned above, the vector fields $B_i$ are non-dynamical dof. We can therefore compute the equation of motion with respect to $B_i$ and integrate them out. By doing so we obtain 
for the non-dynamical degree $B_i$
\begin{equation}
B_{i,\vec{k}} = \frac{a}{2\,m_V^2}\,\frac{\dot{E}_{i,\vec{k}}}{N}\,,
\end{equation}
We can plug this back into the action \ref{perturbed_vector}, which results in
\begin{equation}
S^{(2)}_{\rm vector} = \frac{\mpl^2}{16}\int d^3k\,dt\,k^2a^3 \left[m_V^2
\dot{E}_{i,\vec{k}}^\star\dot{E}^i_{\vec{k}}-m_{T}^2 E_{i,\vec{k}}^\star E^i_{\vec{k}} 
\right]\,.
\end{equation}
In order to avoid ghost instability we have to require that the kinetic term has the right sign. This is guaranteed if we impose $m_V^2 >0$. Assuming that the effective contribution of the mass term and the $\chi$ field on the expansion is not equivalent to something exotic, it is reasonable to expect $\dot{H}<0$. In that case, from Eq.(\ref{eq:veckin}), this condition can be satisfied if $\beta >0$. Similarly as for the tensor perturbations the absence of gradient and tachyonic instability requires $m_T^2>0$.

\subsection{Scalar perturbations}
Scalar fields have been traditionally used as promising candidates for an alternative to the cosmological constant to explain the accelerated expansion of the universe. They naturally appear in gravitational theories beyond GR and high energy physics. Moreover, their existence does not break isotropy. In massive gravity, the helicity-0 degree of freedom of the massive graviton corresponds to a scalar field which might bring along interesting phenomenology. Here we will be concentrating on the stability of the scalar perturbations in our massive gravity model with the effective coupling. As we mentioned above, five dof appear in form of a scalar field $\Phi$, $B$, $\psi$, $E$, $\delta\chi$, from which two ($\Phi$ and $B$) are non-dynamical and will be integrated out. We first calculate the action quadratic in scalar perturbations and introduce their Fourier modes
\begin{eqnarray}
\Phi=\int \frac{d^3k}{(2\pi)^{3/2}} \Phi_{\vec{k}}\;(t)e^{i\vec{k}\cdot\vec{x}} +c.c, && \;\;\;\;\;\;\;\;\;\;\;     B=\int \frac{d^3k}{(2\pi)^{3/2}} B_{\vec{k}}\;(t)e^{i\vec{k}\cdot\vec{x}} +c.c. \nonumber\\
\psi=\int \frac{d^3k}{(2\pi)^{3/2}} \psi_{\vec{k}}\;(t)e^{i\vec{k}\cdot\vec{x}} +c.c, &&\;\;\;\;\;\;\;\;\;\;\;     E=\int \frac{d^3k}{(2\pi)^{3/2}} E_{\vec{k}}\;(t)e^{i\vec{k}\cdot\vec{x}} +c.c.  \nonumber\\
\delta\chi=\int \frac{d^3k}{(2\pi)^{3/2}} && \delta\chi_{\vec{k}}\;(t)e^{i\vec{k}\cdot\vec{x}} +c.c
\label{fourierPhiandpsi}
\end{eqnarray}
First of all, the scalar fields $\Phi$ and $B$ do not carry any time derivatives on them. Hence, we can compute the equations of motion with respect to $\Phi$ 
\begin{eqnarray}
&&H\left(\frac{k^2B_{\vec{k}}}{a}-3\,H\,\Phi_{\vec{k}}+\frac{3\,\dot{\psi}_{\vec{k}}}{N}\right)-\frac{\dot{H}}{N}\left(\frac{\alpha\,\Phi_{\vec{k}}\,N}{N_\eff}+\frac{3\,\beta\,a\,A\,\psi_{\vec{k}}}{a_\eff}\right) \nonumber\\
&&+\frac{k^2}{a^2}\left(\psi_{\vec{k}}+\frac{k^2}{6}E_{\vec{k}}\right)-\frac{\alpha\,a_\eff^3}{2\,\mpl\,a^3}\left(V_{,\chi}\,\delta\chi_{\vec{k}}+\frac{\dot{\chi}_0\,\delta\dot{\chi}_{\vec{k}}}{N_\eff^2}\right)=0\,,
\nonumber\\
\end{eqnarray}
and similarly with respect to $B$
\begin{eqnarray}
\frac{1}{N}\left(\dot{\psi}_{\vec{k}}+\frac{k^2}{6}\,\dot{E}_{\vec{k}}\right)+\frac{\beta\,a^2\,A\,\dot{H}}{(r+1)\,a_\eff\,N}\,\left(1+\frac{a\,N_\eff}{(r+1)\,a_\eff N}\right)\,B_{\vec{k}}  \nonumber\\
-H\,\Phi_{\vec{k}}+\frac{\alpha\,r\,a_\eff^2\,\dot{\chi}_0}{2\,\mpl\,a^2\,N_\eff(r+1)}\,\left(1+\frac{a\,N_\eff}{r\,a_\eff\,N}\right)\delta\chi_{\vec{k}}=0
\,.\nonumber\\
\label{eq:solPhiB}
\end{eqnarray}
and solve them for $B$ and $\Phi$. After plugging back the solutions for $B$ and $\Phi$ the resulting action depends only on the remaining three scalar fields $\psi$, $E$ and $\delta\chi$. 
As we mentioned above, in \cite{deRham:2014naa} it has been explicitly shown that the Boulware-Deser ghost is absent in these exact FLRW solutions, hence we should be able to integrate out one more degree of freedom, which would otherwise correspond to the Boulware-Deser ghost. This becomes manifest after performing the field redefinition
\begin{equation}
\pi_{1,\vec{k}} \equiv \frac{\mpl\,H\,N_\eff}{\alpha\,\dot{\chi}}\delta\chi_{\vec{k}}-\psi_{\vec{k}}\,,\qquad
\pi_{2,\vec{k}} \equiv \frac{E_{\vec{k}}}{k^2}\,.
\label{eq:scalarbasis}
\end{equation}
In term of these new field variables, the remaining degree $\psi$ becomes non-dynamical and can be integrated out using its equation of motion. Furthermore, the $\chi$ dependence is completely removed by the normalization of the $\pi_1$ field. The resulting action can be expressed in the following form
\begin{equation}
S^{(2)}_{\rm scalar}= \frac{\mpl^2}{2}\int d^3k \,dt\,a^3 \,\left(\dot{\Pi}^\dagger\,\hat{K}\,\dot{\Pi} + \dot{\Pi}^\dagger\,\hat{{\cal N}}\,\Pi- \Pi^\dagger\,\hat{{\cal N}}\,\dot{\Pi}-\Pi^\dagger\,\hat{M} \,\Pi\right)\,,
\end{equation}
where $\Pi$ denotes $\Pi=\{ \pi_{1,\vec{k}}, \pi_{2,\vec{k}}\}$and $\hat{K}$, $\hat{M}$ and $\hat{{\cal N}}$ are $2\times2$ real, time-dependent matrices with the properties $\hat{K}^T=\hat{K}$, $\hat{M}^T=\hat{M}$ and $\hat{{\cal N}}^T=-\hat{{\cal N}}$. In terms of the new variables (\ref{eq:scalarbasis}) we have non-vanishing off-diagonal components, which are not suitable for presentation. Nevertheless, we can diagonalize the kinetic term through the following field redefinition
\begin{equation}
\hat{Z} = \hat{R}^{-1}\,\Pi\,,
\end{equation}
with the rotation matrix
\begin{equation}
\hat{R} = \left(
\begin{array}{cc}
1 & 0  \\
-\frac{\hat{K}_{12}}{\hat{K}_{22}} & 1
\end{array}
\right)\,.
\end{equation}
In this new basis the kinetic matrix becomes diagonal
\begin{equation}
\hat{R}^T\,\hat{K}\,\hat{R} = {\rm diag}\left(\frac{\det \hat{K}}{\hat{K}_{22}}\,,\;\hat{K}_{22}\right)\,.
\end{equation}
Despite the cumbersome mathematical expressions involved, we can further simplify $\det \hat{K}$ and $\det \hat{K}/\hat{K}_{22}$ and impose from their positivity also the positivity of $\hat{K}_{22}$
\begin{eqnarray}
\frac{\det \hat{K}}{\hat{K}_{22}} &=& -6\left[1+\frac{3 H^2\,N_\eff}{\alpha\,\dot{H}}
-\frac{1}{3\,\beta\,A\,{\cal Y}_1}\left(
\frac{{\cal Y}_2^2\,(r-1)N_\eff}{N{\cal Y}_3} - \frac{k^2(1+r)^2a_\eff^2N}{a^4\,\dot{H}}\right)
\right]^{-1}\,,\nonumber\\
\det \hat{K}&=&-\Bigg[
1+
\frac{3\,H^2\,N_\eff}{\alpha\,\dot{H}}
-\frac{N_\eff\,(r-1)}{3\,\beta\,{\cal Y}_1 \,{\cal Y}_3\,A\,N}
\left({\cal Y}_2^2 - \frac{k^4a_\eff^2N_\eff^2}{a^6H^4N^2}
\left(\frac{3\,H^2\,N_\eff}{\alpha\,\dot{H}}-1\right)
\right)\nonumber\\
&&\qquad\qquad\qquad\qquad
+\frac{k^2(1+r)^2a_\eff^2N_\eff}{{\cal Y}_3\,a^4\dot{H}}\left(9(r-1)-\frac{{\cal Y}_3\,H^2\,N}{\alpha\,\beta\,A\,{\cal Y}_1\,\dot{H}}\right)\Bigg]^{-1}\,,
\end{eqnarray}
where we introduced the following definitions for the sake of clarity of the cumbersome expressions
\begin{eqnarray}
{\cal Y}_1 &\equiv& \frac{N_\eff}{N}+\frac{a_\eff}{a}(1+r)\,,\nonumber\\
{\cal Y}_2 &\equiv& \frac{a_\eff N_\eff}{a\,N}\frac{k^2}{a^2\,H^2}+9\,\beta\,A\,{\cal Y}_1\,,\nonumber\\
{\cal Y}_3 &\equiv&\frac{(r-1)\dot{H}\,a_\eff^2 N_\eff}{H^2 a^2\,N^2}\left(\frac{k^2}{a^2H^2}+\frac{9\,\alpha\,\beta \,a^2 N\,A\,{\cal Y}_1}{a_\eff^2\,N_\eff}\right)+3\,{\cal Y}_1\frac{m_{T}^2\,a_\eff^2}{H^2a^2}\,.
\end{eqnarray}
In the following we will investigate the positivity of the kinetic terms in the UV and IR regime. First of all, in the UV the two kinetic terms are
\begin{equation}
\frac{\det \hat{K}}{\hat{K}_{22}}= -\frac{18\,\beta\,a^2A\,\dot{H}}{N\,k^2(\alpha\,r+\beta\,A)}+{\cal O}(k^{-4})
\,,
\qquad
\hat{K}_{22}= -\frac{\alpha\,\dot{H}}{18\,H^2\,N_\eff} + {\cal O}(k^{-2})\,.
\end{equation}
By construction, the scale factors and the lapses of $f$ and $g$ metrics are positive. Thus, in order to have both kinetic terms simultaneously positive at high energies, we need to impose $\alpha>0$ and $\beta>0$. On the other hand, at low energies we have
\begin{equation}
\frac{\det \hat{K}}{\hat{K}_{22}}= 6\left[-1+\left(1+\frac{\alpha\,\dot{H}}{3\,N_\eff H^2}+\frac{\alpha^2\beta \,a^2(r-1)A\,\dot{H}^2}{a_\eff^2H^2N\,N_\eff\,m_{T}^2}\right)^{-1}\right]+{\cal O}(k^2)
\,,
\qquad
\hat{K}_{22}= \frac{1}{6}+{\cal O}(k^2)\,.
\end{equation}
Clearly, the second mode is not a ghost at low energies. The first mode however imposes some condition. Assuming $\alpha>0$, $\beta>0$, $m_{T}^2>0$ and $\dot{H}<0$, we find an upper bound on $r$, given by
\begin{equation}
r < 1 +\frac{m_{T}^2\,a_\eff^2}{3\,\alpha\,\beta\,a^2\,A \left(-\frac{\dot{H}}{N}\right)}\,.
\end{equation}

\section{Conclusions}
\label{sec:conclusion}

In massive (bi-) gravity, the two 
metrics are put on equal footing and therefore it is a natural question
to ask how the matter fields would couple to these two metrics. One
immediate possibility would be to minimally couple each metric to its own separate
matter sector. This form of coupling is free of the Boulware-Deser ghost
at all scales and maintains this property also at the quantum level. The
integration over matter loops generates quantum corrections in form of
two cosmological constants for each metric respectively. Another obvious
possibility would be to couple the matter sector to both metrics
simultaneously. Nevertheless 
 a generic coupling of this type reintroduces the
Boulware-Deser ghost already at the classical level and the quantum 
corrections detune the potential structure with a 
scale arbitrarily below the strong coupling scale. A third
possibility consists of the first type of coupling but with an interaction between
the different matter sectors. However, the quantum corrections would
again yield a detuning with a resulting Boulware-Deser ghost at an
unacceptable low scale. Similarly one could also try to couple parts of
the matter sector to two different metrics, for instance the kinetic
term would couple to one metric whereas the potential to the other
metric. However, it has been shown that also in this case the coupling
has disastrous effects at the quantum level and consequently destroying
the validity of the classical theory. Last but not least, another
promising coupling between the matter sector and an effective metric
composed of the two metric was proposed and it was shown
that in this case the theory is free of the Boulware-Deser
ghost below the strong coupling scale at both classical and
quantum levels.\\ 

The presence of this new effective metric opens up new directions of
investigation for phenomenology. First of all, the no-go theorem for
flat FLRW solutions can be circumvented with this composite metric. This
work goes along this line and was dedicated to the analysis of
cosmological perturbations in the context of non-linear massive gravity
with this very specific effective coupling to the matter fields. After
studying in detail the background evolution for the flat 
FLRW, we considered perturbations on top of this background. We studied
the stability of tensor, vector and scalar perturbations and put
constraints on the parameters of the theory coming from the requirement
of absence of ghost and gradient instabilities. 
Because of the non-derivative nature of the dRGT mass term, the kinetic and gradient terms of the tensor perturbations are unchanged with respect to GR.
We commented on the possibility of a
tachyonic instability for $m_T^2<0$ and estimated the time scale within
which the instability would be negligible. Concerning the vector
perturbations, we first had to integrate out the non-dynamical vector
modes, after which the kinetic term obtained the correct sign upon
imposing the conditions $\dot{H}<0$ and $\beta >0$. The absence of
tachyonic instability requires $m_T^2>0$. The analysis of the scalar
perturbations is more involved. Out of the five scalar 
variables
we first eliminated the two non-dynamical fields $\Phi$ and
$B$. In a more suitable basis of fields we were also able to integrate
out the would-be Boulware-Deser ghost and obtained the final action for
the two dynamical scalar fields. We studied the kinetic terms of the two
remaining scalar fields in the UV and IR regime and the correct sign for
the kinetic terms requires $\alpha>0$ and $\beta>0$ for the two
parameters in the effective matter coupling. \\

The new matter coupling evades not only the no-go result
for the flat FLRW background but also yet another no-go found in
\cite{PhysRevLett.109.171101}. While the former no-go could be
circumvented by self-accelerating open FLRW
solutions~\cite{Gumrukcuoglu:2011ew},
the latter no-go tells that perturbations around the self-accelerating
background exhibit ghost instability at nonlinear 
order, if the matter coupling is minimal. This is due to the fact that
the quadratic kinetic terms of three among five physical degrees of
freedom are proportional to the background St\"uckelberg equation of
motion, thus vanish~\cite{Gumrukcuoglu:2011zh} and that they acquire
non-vanishing kinetic terms only at cubic order \cite{PhysRevD.86.124019}. On the other hand, with
the new matter coupling, all five physical degrees of freedom have
non-vanishing kinetic terms already at the quadratic order, on both flat
and open FLRW backgrounds. In this way, the new matter coupling helps
evading the two previous no-go results that have been obstacles for
massive gravity cosmology.

\acknowledgments

We would like to thank N.~Tanahashi for very useful discussions.
AEG acknowledges financial support from the European Research
Council under the European Union's Seventh Framework Programme
(FP7/2007-2013) / ERC Grant Agreement n. 306425 ``Challenging General
Relativity''.  
This work was supported in part by WPI Initiative, MEXT, Japan and
Grant-in-Aid for Scientific Research 24540256. 

\appendix

	\bibliographystyle{JHEPmodplain}
	\bibliography{EffectiveCoupling_cosmology}

\providecommand{\href}[2]{#2}\begingroup\raggedright\begin{thebibliography}{10}

\bibitem{Weinberg:1988cp}
S.~Weinberg, {\it {The cosmological constant problem}},  {\sl Rev.Mod.Phys.}
  {\bf 61} (1989) 1--23.

\bibitem{Perlmutter:1997zf}
{\bf Supernova Cosmology Project} Collaboration, S.~Perlmutter {\em et~al.},
  {\it {Discovery of a supernova explosion at half the age of the Universe and
  its cosmological implications}},  {\sl Nature} {\bf 391} (1998) 51--54,
  [\href{http://arxiv.org/abs/astro-ph/9712212}{{\sf arXiv:astro-ph/9712212}}].

\bibitem{Riess:1998cb}
{\bf Supernova Search Team} Collaboration, A.~G. Riess {\em et~al.}, {\it
  {Observational evidence from supernovae for an accelerating universe and a
  cosmological constant}},  {\sl Astron. J.} {\bf 116} (1998) 1009--1038,
  [\href{http://arxiv.org/abs/astro-ph/9805201}{{\sf arXiv:astro-ph/9805201}}].

\bibitem{Tonry:2003zg}
{\bf Supernova Search Team} Collaboration, J.~L. Tonry {\em et~al.}, {\it
  {Cosmological results from high-z supernovae}},  {\sl Astrophys. J.} {\bf
  594} (2003) 1--24, [\href{http://arxiv.org/abs/astro-ph/0305008}{{\sf
  arXiv:astro-ph/0305008}}].

\bibitem{Horava:2009uw}
P.~Horava, {\it {Quantum Gravity at a Lifshitz Point}},  {\sl Phys.Rev.} {\bf
  D79} (2009) 084008, [\href{http://arxiv.org/abs/0901.3775}{{\sf
  arXiv:0901.3775}}], [\href{http://dx.doi.org/10.1103/PhysRevD.79.084008}{{\sf
  doi:10.1103/PhysRevD.79.084008}}].

\bibitem{Blas:2009yd}
D.~Blas, O.~Pujolas, and S.~Sibiryakov, {\it {On the Extra Mode and
  Inconsistency of Horava Gravity}},  {\sl JHEP} {\bf 0910} (2009) 029,
  [\href{http://arxiv.org/abs/0906.3046}{{\sf arXiv:0906.3046}}],
  [\href{http://dx.doi.org/10.1088/1126-6708/2009/10/029}{{\sf
  doi:10.1088/1126-6708/2009/10/029}}].

\bibitem{Blas:2009qj}
D.~Blas, O.~Pujolas, and S.~Sibiryakov, {\it {Consistent Extension of Horava
  Gravity}},  {\sl Phys.Rev.Lett.} {\bf 104} (2010) 181302,
  [\href{http://arxiv.org/abs/0909.3525}{{\sf arXiv:0909.3525}}],
  [\href{http://dx.doi.org/10.1103/PhysRevLett.104.181302}{{\sf
  doi:10.1103/PhysRevLett.104.181302}}].

\bibitem{Mukohyama:2009zs}
S.~Mukohyama, K.~Nakayama, F.~Takahashi, and S.~Yokoyama, {\it
  {Phenomenological Aspects of Horava-Lifshitz Cosmology}},  {\sl Phys.Lett.}
  {\bf B679} (2009) 6--9, [\href{http://arxiv.org/abs/0905.0055}{{\sf
  arXiv:0905.0055}}],
  [\href{http://dx.doi.org/10.1016/j.physletb.2009.07.005}{{\sf
  doi:10.1016/j.physletb.2009.07.005}}].

\bibitem{Blas:2010hb}
D.~Blas, O.~Pujolas, and S.~Sibiryakov, {\it {Models of non-relativistic
  quantum gravity: The Good, the bad and the healthy}},  {\sl JHEP} {\bf 1104}
  (2011) 018, [\href{http://arxiv.org/abs/1007.3503}{{\sf arXiv:1007.3503}}],
  [\href{http://dx.doi.org/10.1007/JHEP04(2011)018}{{\sf
  doi:10.1007/JHEP04(2011)018}}].

\bibitem{Nojiri:2010wj}
S.~Nojiri and S.~D. Odintsov, {\it {Unified cosmic history in modified gravity:
  from F(R) theory to Lorentz non-invariant models}},  {\sl Phys.Rept.} {\bf
  505} (2011) 59--144, [\href{http://arxiv.org/abs/1011.0544}{{\sf
  arXiv:1011.0544}}],
  [\href{http://dx.doi.org/10.1016/j.physrep.2011.04.001}{{\sf
  doi:10.1016/j.physrep.2011.04.001}}].

\bibitem{Mukohyama:2010xz}
S.~Mukohyama, {\it {Horava-Lifshitz Cosmology: A Review}},  {\sl
  Class.Quant.Grav.} {\bf 27} (2010) 223101,
  [\href{http://arxiv.org/abs/1007.5199}{{\sf arXiv:1007.5199}}],
  [\href{http://dx.doi.org/10.1088/0264-9381/27/22/223101}{{\sf
  doi:10.1088/0264-9381/27/22/223101}}].

\bibitem{Deser:1998rj}
S.~Deser and G.~Gibbons, {\it {Born-Infeld-Einstein actions?}},  {\sl
  Class.Quant.Grav.} {\bf 15} (1998) L35--L39,
  [\href{http://arxiv.org/abs/hep-th/9803049}{{\sf arXiv:hep-th/9803049}}],
  [\href{http://dx.doi.org/10.1088/0264-9381/15/5/001}{{\sf
  doi:10.1088/0264-9381/15/5/001}}].

\bibitem{Ketov:2001dq}
S.~V. Ketov, {\it {Many faces of Born-Infeld theory}},
  \href{http://arxiv.org/abs/hep-th/0108189}{{\sf arXiv:hep-th/0108189}}.

\bibitem{Wohlfarth:2003ss}
M.~N. Wohlfarth, {\it {Gravity a la Born-Infeld}},  {\sl Class.Quant.Grav.}
  {\bf 21} (2004) 1927, [\href{http://arxiv.org/abs/hep-th/0310067}{{\sf
  arXiv:hep-th/0310067}}],
  [\href{http://dx.doi.org/10.1088/0264-9381/21/8/001}{{\sf
  doi:10.1088/0264-9381/21/8/001}}].

\bibitem{Pani:2012qb}
P.~Pani, T.~Delsate, and V.~Cardoso, {\it {Eddington-inspired Born-Infeld
  gravity. Phenomenology of non-linear gravity-matter coupling}},  {\sl
  Phys.Rev.} {\bf D85} (2012) 084020,
  [\href{http://arxiv.org/abs/1201.2814}{{\sf arXiv:1201.2814}}],
  [\href{http://dx.doi.org/10.1103/PhysRevD.85.084020}{{\sf
  doi:10.1103/PhysRevD.85.084020}}].

\bibitem{Scargill:2012kg}
J.~H. Scargill, M.~Banados, and P.~G. Ferreira, {\it {Cosmology with
  Eddington-inspired Gravity}},  {\sl Phys.Rev.} {\bf D86} (2012) 103533,
  [\href{http://arxiv.org/abs/1210.1521}{{\sf arXiv:1210.1521}}],
  [\href{http://dx.doi.org/10.1103/PhysRevD.86.103533}{{\sf
  doi:10.1103/PhysRevD.86.103533}}].

\bibitem{Olmo:2013gqa}
G.~J. Olmo, D.~Rubiera-Garcia, and H.~Sanchis-Alepuz, {\it {Geonic black holes
  and remnants in Eddington-inspired Born-Infeld gravity}},  {\sl Eur.Phys.J.}
  {\bf C74} (2014) 2804, [\href{http://arxiv.org/abs/1311.0815}{{\sf
  arXiv:1311.0815}}],
  [\href{http://dx.doi.org/10.1140/epjc/s10052-014-2804-8}{{\sf
  doi:10.1140/epjc/s10052-014-2804-8}}].

\bibitem{Odintsov:2014yaa}
S.~D. Odintsov, G.~J. Olmo, and D.~Rubiera-Garcia, {\it {Born-Infeld gravity
  and its functional extensions}},  {\sl Phys.Rev.} {\bf D90} (2014) 044003,
  [\href{http://arxiv.org/abs/1406.1205}{{\sf arXiv:1406.1205}}],
  [\href{http://dx.doi.org/10.1103/PhysRevD.90.044003}{{\sf
  doi:10.1103/PhysRevD.90.044003}}].

\bibitem{Komada:2014asa}
M.~Komada and S.~Nojiri, {\it {Palatini-Born-Infeld Gravity and Black Hole
  Formation}},  \href{http://arxiv.org/abs/1409.1663}{{\sf arXiv:1409.1663}}.

\bibitem{Jimenez:2014fla}
J.~B. Jim{\'e}nez, L.~Heisenberg, and G.~J. Olmo, {\it {Infrared lessons for
  ultraviolet gravity: the case of massive gravity and Born-Infeld}},
  \href{http://arxiv.org/abs/1409.0233}{{\sf arXiv:1409.0233}}.

\bibitem{Vollick:2005gc}
D.~N. Vollick, {\it {Born-Infeld-Einstein theory with matter}},  {\sl
  Phys.Rev.} {\bf D72} (2005) 084026,
  [\href{http://arxiv.org/abs/gr-qc/0506091}{{\sf arXiv:gr-qc/0506091}}],
  [\href{http://dx.doi.org/10.1103/PhysRevD.72.084026}{{\sf
  doi:10.1103/PhysRevD.72.084026}}].

\bibitem{Pani:2012qd}
P.~Pani and T.~P. Sotiriou, {\it {Surface singularities in Eddington-inspired
  Born-Infeld gravity}},  {\sl Phys.Rev.Lett.} {\bf 109} (2012) 251102,
  [\href{http://arxiv.org/abs/1209.2972}{{\sf arXiv:1209.2972}}],
  [\href{http://dx.doi.org/10.1103/PhysRevLett.109.251102}{{\sf
  doi:10.1103/PhysRevLett.109.251102}}].

\bibitem{deRham:2010ik}
C.~de~Rham and G.~Gabadadze, {\it {Generalization of the Fierz-Pauli action}},
  {\sl Phys.Rev.} {\bf D82} (2010) 044020,
  [\href{http://arxiv.org/abs/1007.0443}{{\sf arXiv:1007.0443}}].

\bibitem{deRham:2010kj}
C.~de~Rham, G.~Gabadadze, and A.~J. Tolley, {\it {Resummation of massive
  gravity}},  {\sl Phys.Rev.Lett.} {\bf 106} (2011) 231101,
  [\href{http://arxiv.org/abs/1011.1232}{{\sf arXiv:1011.1232}}].

\bibitem{Fierz:1939ix}
M.~Fierz and W.~Pauli, {\it {On relativistic wave equations for particles of
  arbitrary spin in an electromagnetic field}},  {\sl Proc.Roy.Soc.Lond.} {\bf
  A173} (1939) 211--232.

\bibitem{VanNieuwenhuizen:1973fi}
P.~Van~Nieuwenhuizen, {\it {On ghost-free tensor lagrangians and linearized
  gravitation}},  {\sl Nucl.Phys.} {\bf B60} (1973) 478--492.

\bibitem{Hinterbichler:2011tt}
K.~Hinterbichler, {\it {Theoretical Aspects of Massive Gravity}},  {\sl
  Rev.Mod.Phys.} {\bf 84} (2012) 671--710,
  [\href{http://arxiv.org/abs/1105.3735}{{\sf arXiv:1105.3735}}],
  [\href{http://dx.doi.org/10.1103/RevModPhys.84.671}{{\sf
  doi:10.1103/RevModPhys.84.671}}].

\bibitem{deRham:2014zqa}
C.~de~Rham, {\it {Massive Gravity}},  {\sl Living Rev.Rel.} {\bf 17} (2014) 7,
  [\href{http://arxiv.org/abs/1401.4173}{{\sf arXiv:1401.4173}}],
  [\href{http://dx.doi.org/10.12942/lrr-2014-7}{{\sf
  doi:10.12942/lrr-2014-7}}].

\bibitem{Boulware:1973my}
D.~Boulware and S.~Deser, {\it {Can gravitation have a finite range?}},  {\sl
  Phys.Rev.} {\bf D6} (1972) 3368--3382.

\bibitem{Hassan:2011hr}
S.~Hassan and R.~A. Rosen, {\it {Resolving the ghost problem in non-linear
  massive gravity}},  {\sl Phys.Rev.Lett.} {\bf 108} (2012) 041101,
  [\href{http://arxiv.org/abs/1106.3344}{{\sf arXiv:1106.3344}}].

\bibitem{deRham:2011rn}
C.~de~Rham, G.~Gabadadze, and A.~J. Tolley, {\it {Ghost free massive gravity in
  the St\'uckelberg language}},  {\sl Phys.Lett.} {\bf B711} (2012) 190--195,
  [\href{http://arxiv.org/abs/1107.3820}{{\sf arXiv:1107.3820}}].

\bibitem{Hassan:2011ea}
S.~Hassan and R.~A. Rosen, {\it {Confirmation of the secondary constraint and
  absence of ghost in massive gravity and bimetric gravity}},  {\sl JHEP} {\bf
  1204} (2012) 123, [\href{http://arxiv.org/abs/1111.2070}{{\sf
  arXiv:1111.2070}}].

\bibitem{deRham:2011qq}
C.~de~Rham, G.~Gabadadze, and A.~J. Tolley, {\it {Helicity decomposition of
  ghost-free massive gravity}},  {\sl JHEP} {\bf 1111} (2011) 093,
  [\href{http://arxiv.org/abs/1108.4521}{{\sf arXiv:1108.4521}}].

\bibitem{Mirbabayi:2011aa}
M.~Mirbabayi, {\it {A proof of ghost freedom in de Rham-Gabadadze-Tolley
  massive gravity}},  {\sl Phys.Rev.} {\bf D86} (2012) 084006,
  [\href{http://arxiv.org/abs/1112.1435}{{\sf arXiv:1112.1435}}].

\bibitem{Golovnev:2011aa}
A.~Golovnev, {\it {On the Hamiltonian analysis of non-linear massive gravity}},
   {\sl Phys.Lett.} {\bf B707} (2012) 404--408,
  [\href{http://arxiv.org/abs/1112.2134}{{\sf arXiv:1112.2134}}].

\bibitem{Hassan:2012qv}
S.~Hassan, A.~Schmidt-May, and M.~von Strauss, {\it {Proof of consistency of
  nonlinear massive gravity in the St\'uckelberg formulation}},  {\sl
  Phys.Lett.} {\bf B715} (2012) 335--339,
  [\href{http://arxiv.org/abs/1203.5283}{{\sf arXiv:1203.5283}}].

\bibitem{Hinterbichler:2012cn}
K.~Hinterbichler and R.~A. Rosen, {\it {Interacting Spin-2 Fields}},  {\sl
  JHEP} {\bf 1207} (2012) 047, [\href{http://arxiv.org/abs/1203.5783}{{\sf
  arXiv:1203.5783}}].

\bibitem{Kluson:2012wf}
J.~Kluson, {\it {Non-linear massive gravity with additional primary constraint
  and absence of ghosts}},  {\sl Phys.Rev.} {\bf D86} (2012) 044024,
  [\href{http://arxiv.org/abs/1204.2957}{{\sf arXiv:1204.2957}}].

\bibitem{Deffayet:2012zc}
C.~Deffayet, J.~Mourad, and G.~Zahariade, {\it {A note on 'symmetric' vielbeins
  in bimetric, massive, perturbative and non perturbative gravities}},  {\sl
  JHEP} {\bf 1303} (2013) 086, [\href{http://arxiv.org/abs/1208.4493}{{\sf
  arXiv:1208.4493}}].

\bibitem{Nicolis:2008in}
A.~Nicolis, R.~Rattazzi, and E.~Trincherini, {\it {The galileon as a local
  modification of gravity}},  {\sl Phys.Rev.} {\bf D79} (2009) 064036,
  [\href{http://arxiv.org/abs/0811.2197}{{\sf arXiv:0811.2197}}].

\bibitem{deRham:2010tw}
C.~de~Rham, G.~Gabadadze, L.~Heisenberg, and D.~Pirtskhalava, {\it {Cosmic
  acceleration and the helicity-0 graviton}},  {\sl Phys.Rev.} {\bf D83} (2011)
  103516, [\href{http://arxiv.org/abs/1010.1780}{{\sf arXiv:1010.1780}}].

\bibitem{Ondo:2013wka}
N.~A. Ondo and A.~J. Tolley, {\it {Complete decoupling limit of ghost-free
  massive gravity}},  \href{http://arxiv.org/abs/1307.4769}{{\sf
  arXiv:1307.4769}}.

\bibitem{deRham:2012ew}
C.~de~Rham, G.~Gabadadze, L.~Heisenberg, and D.~Pirtskhalava, {\it
  {Non-renormalization and naturalness in a class of scalar-tensor theories}},
  {\sl Phys.Rev.} {\bf D87} (2012) [\href{http://arxiv.org/abs/1212.4128}{{\sf
  arXiv:1212.4128}}].

\bibitem{Brouzakis:2013lla}
N.~Brouzakis, A.~Codello, N.~Tetradis, and O.~Zanusso, {\it {Quantum
  corrections in Galileon theories}},  {\sl Phys.Rev.} {\bf D89} (2014) 125017,
  [\href{http://arxiv.org/abs/1310.0187}{{\sf arXiv:1310.0187}}],
  [\href{http://dx.doi.org/10.1103/PhysRevD.89.125017}{{\sf
  doi:10.1103/PhysRevD.89.125017}}].

\bibitem{deRham:2014wfa}
C.~de~Rham and R.~H. Ribeiro, {\it {Riding on irrelevant operators}},
  \href{http://arxiv.org/abs/1405.5213}{{\sf arXiv:1405.5213}}.

\bibitem{Heisenberg:2014raa}
L.~Heisenberg, {\it {Quantum corrections in Galileons from matter loops}},
  \href{http://arxiv.org/abs/1408.0267}{{\sf arXiv:1408.0267}}.

\bibitem{Park:2010rp}
M.~Park, {\it {Quantum aspects of massive gravity}},  {\sl Class.Quant.Grav.}
  {\bf 28} (2011) 105012, [\href{http://arxiv.org/abs/1009.4369}{{\sf
  arXiv:1009.4369}}].

\bibitem{Buchbinder:2012wb}
I.~Buchbinder, D.~Pereira, and I.~Shapiro, {\it {One-loop divergences in
  massive gravity theory}},  {\sl Phys.Lett.} {\bf B712} (2012) 104--108,
  [\href{http://arxiv.org/abs/1201.3145}{{\sf arXiv:1201.3145}}],
  [\href{http://dx.doi.org/10.1016/j.physletb.2012.04.045}{{\sf
  doi:10.1016/j.physletb.2012.04.045}}].

\bibitem{deRham:2013qqa}
C.~de~Rham, L.~Heisenberg, and R.~H. Ribeiro, {\it {Quantum Corrections in
  Massive Gravity}},  {\sl Phys.Rev.} {\bf D88} (2013) 084058,
  [\href{http://arxiv.org/abs/1307.7169}{{\sf arXiv:1307.7169}}],
  [\href{http://dx.doi.org/10.1103/PhysRevD.88.084058}{{\sf
  doi:10.1103/PhysRevD.88.084058}}].

\bibitem{deRham:2014naa}
C.~de~Rham, L.~Heisenberg, and R.~H. Ribeiro, {\it {On couplings to matter in
  massive (bi-)gravity}},  \href{http://arxiv.org/abs/1408.1678}{{\sf
  arXiv:1408.1678}}.

\bibitem{PhysRevD.84.124046}
G.~D'Amico, C.~de~Rham, S.~Dubovsky, G.~Gabadadze, D.~Pirtskhalava, and A.~J.
  Tolley, {\it Massive cosmologies},  {\sl Phys. Rev. D} {\bf 84} (Dec, 2011)
  124046, [\href{http://dx.doi.org/10.1103/PhysRevD.84.124046}{{\sf
  doi:10.1103/PhysRevD.84.124046}}].

\bibitem{PhysRevLett.109.171101}
A.~De~Felice, A.~E. G\"umr\"uk\ifmmode \mbox{\c{c}}\else
  \c{c}\fi{}\"uo\ifmmode~\breve{g}\else \u{g}\fi{}lu, and S.~Mukohyama, {\it
  Massive gravity: Nonlinear instability of a homogeneous and isotropic
  universe},  {\sl Phys. Rev. Lett.} {\bf 109} (Oct, 2012) 171101,
  [\href{http://dx.doi.org/10.1103/PhysRevLett.109.171101}{{\sf
  doi:10.1103/PhysRevLett.109.171101}}].

\bibitem{deRham:2011by}
C.~de~Rham and L.~Heisenberg, {\it {Cosmology of the Galileon from Massive
  Gravity}},  {\sl Phys.Rev.} {\bf D84} (2011) 043503,
  [\href{http://arxiv.org/abs/1106.3312}{{\sf arXiv:1106.3312}}],
  [\href{http://dx.doi.org/10.1103/PhysRevD.84.043503}{{\sf
  doi:10.1103/PhysRevD.84.043503}}].

\bibitem{Chamseddine:2011bu}
A.~H. Chamseddine and M.~S. Volkov, {\it {Cosmological solutions with massive
  gravitons}},  {\sl Phys.Lett.} {\bf B704} (2011) 652--654,
  [\href{http://arxiv.org/abs/1107.5504}{{\sf arXiv:1107.5504}}].

\bibitem{Koyama:2011xz}
K.~Koyama, G.~Niz, and G.~Tasinato, {\it {Analytic solutions in non-linear
  massive gravity}},  {\sl Phys.Rev.Lett.} {\bf 107} (2011) 131101,
  [\href{http://arxiv.org/abs/1103.4708}{{\sf arXiv:1103.4708}}],
  [\href{http://dx.doi.org/10.1103/PhysRevLett.107.131101}{{\sf
  doi:10.1103/PhysRevLett.107.131101}}].

\bibitem{Koyama:2011wx}
K.~Koyama, G.~Niz, and G.~Tasinato, {\it {The Self-Accelerating Universe with
  Vectors in Massive Gravity}},  {\sl JHEP} {\bf 1112} (2011) 065,
  [\href{http://arxiv.org/abs/1110.2618}{{\sf arXiv:1110.2618}}],
  [\href{http://dx.doi.org/10.1007/JHEP12(2011)065}{{\sf
  doi:10.1007/JHEP12(2011)065}}].

\bibitem{Gumrukcuoglu:2011zh}
A.~E. Gumrukcuoglu, C.~Lin, and S.~Mukohyama, {\it {Cosmological perturbations
  of self-accelerating universe in nonlinear massive gravity}},  {\sl JCAP}
  {\bf 1203} (2012) 006, [\href{http://arxiv.org/abs/1111.4107}{{\sf
  arXiv:1111.4107}}],
  [\href{http://dx.doi.org/10.1088/1475-7516/2012/03/006}{{\sf
  doi:10.1088/1475-7516/2012/03/006}}].

\bibitem{Gratia:2012wt}
P.~Gratia, W.~Hu, and M.~Wyman, {\it {Self-accelerating Massive Gravity: Exact
  solutions for any isotropic matter distribution}},  {\sl Phys.Rev.} {\bf D86}
  (2012) 061504, [\href{http://arxiv.org/abs/1205.4241}{{\sf
  arXiv:1205.4241}}], [\href{http://dx.doi.org/10.1103/PhysRevD.86.061504}{{\sf
  doi:10.1103/PhysRevD.86.061504}}].

\bibitem{Vakili:2012tm}
B.~Vakili and N.~Khosravi, {\it {Classical and quantum massive cosmology for
  the open FRW universe}},  {\sl Phys.Rev.} {\bf D85} (2012) 083529,
  [\href{http://arxiv.org/abs/1204.1456}{{\sf arXiv:1204.1456}}],
  [\href{http://dx.doi.org/10.1103/PhysRevD.85.083529}{{\sf
  doi:10.1103/PhysRevD.85.083529}}].

\bibitem{Kobayashi:2012fz}
T.~Kobayashi, M.~Siino, M.~Yamaguchi, and D.~Yoshida, {\it {New Cosmological
  Solutions in Massive Gravity}},  {\sl Phys.Rev.} {\bf D86} (2012) 061505,
  [\href{http://arxiv.org/abs/1205.4938}{{\sf arXiv:1205.4938}}],
  [\href{http://dx.doi.org/10.1103/PhysRevD.86.061505}{{\sf
  doi:10.1103/PhysRevD.86.061505}}].

\bibitem{Fasiello:2012rw}
M.~Fasiello and A.~J. Tolley, {\it {Cosmological perturbations in Massive
  Gravity and the Higuchi bound}},  {\sl JCAP} {\bf 1211} (2012) 035,
  [\href{http://arxiv.org/abs/1206.3852}{{\sf arXiv:1206.3852}}],
  [\href{http://dx.doi.org/10.1088/1475-7516/2012/11/035}{{\sf
  doi:10.1088/1475-7516/2012/11/035}}].

\bibitem{Volkov:2012zb}
M.~S. Volkov, {\it {Exact self-accelerating cosmologies in the ghost-free
  massive gravity -- the detailed derivation}},  {\sl Phys.Rev.} {\bf D86}
  (2012) 104022, [\href{http://arxiv.org/abs/1207.3723}{{\sf
  arXiv:1207.3723}}], [\href{http://dx.doi.org/10.1103/PhysRevD.86.104022}{{\sf
  doi:10.1103/PhysRevD.86.104022}}].

\bibitem{Tasinato:2012ze}
G.~Tasinato, K.~Koyama, and G.~Niz, {\it {Vector instabilities and
  self-acceleration in the decoupling limit of massive gravity}},  {\sl
  Phys.Rev.} {\bf D87} (2013) 064029,
  [\href{http://arxiv.org/abs/1210.3627}{{\sf arXiv:1210.3627}}].

\bibitem{DeFelice:2013bxa}
A.~De~Felice, A.~E. G{\"u}mr{\"u}k{\c c}{\"u}o{\u g}lu, C.~Lin, and
  S.~Mukohyama, {\it {On the cosmology of massive gravity}},  {\sl
  Class.Quant.Grav.} {\bf 30} (2013) 184004,
  [\href{http://arxiv.org/abs/1304.0484}{{\sf arXiv:1304.0484}}],
  [\href{http://dx.doi.org/10.1088/0264-9381/30/18/184004}{{\sf
  doi:10.1088/0264-9381/30/18/184004}}].

\bibitem{Fasiello:2013woa}
M.~Fasiello and A.~J. Tolley, {\it {Cosmological Stability Bound in Massive
  Gravity and Bigravity}},  {\sl JCAP} {\bf 1312} (2013) 002,
  [\href{http://arxiv.org/abs/1308.1647}{{\sf arXiv:1308.1647}}],
  [\href{http://dx.doi.org/10.1088/1475-7516/2013/12/002}{{\sf
  doi:10.1088/1475-7516/2013/12/002}}].

\bibitem{Heisenberg:2014kea}
L.~Heisenberg, R.~Kimura, and K.~Yamamoto, {\it {Cosmology of the proxy theory
  to massive gravity}},  {\sl Phys.Rev.} {\bf D89} (2014) 103008,
  [\href{http://arxiv.org/abs/1403.2049}{{\sf arXiv:1403.2049}}],
  [\href{http://dx.doi.org/10.1103/PhysRevD.89.103008}{{\sf
  doi:10.1103/PhysRevD.89.103008}}].

\bibitem{Comelli:2013tja}
D.~Comelli, F.~Nesti, and L.~Pilo, {\it {Cosmology in General Massive Gravity
  Theories}},  {\sl JCAP} {\bf 1405} (2014) 036,
  [\href{http://arxiv.org/abs/1307.8329}{{\sf arXiv:1307.8329}}],
  [\href{http://dx.doi.org/10.1088/1475-7516/2014/05/036}{{\sf
  doi:10.1088/1475-7516/2014/05/036}}].

\bibitem{Motloch:2014nwa}
P.~Motloch and W.~Hu, {\it {Self-accelerating Massive Gravity: Covariant
  Perturbation Theory}},  \href{http://arxiv.org/abs/1409.2204}{{\sf
  arXiv:1409.2204}}.

\bibitem{Hassan:2011vm}
S.~Hassan and R.~A. Rosen, {\it {On Non-Linear Actions for Massive Gravity}},
  {\sl JHEP} {\bf 1107} (2011) 009, [\href{http://arxiv.org/abs/1103.6055}{{\sf
  arXiv:1103.6055}}], [\href{http://dx.doi.org/10.1007/JHEP07(2011)009}{{\sf
  doi:10.1007/JHEP07(2011)009}}].

\bibitem{Hassan:2011zd}
S.~Hassan and R.~A. Rosen, {\it {Bimetric Gravity from Ghost-free Massive
  Gravity}},  {\sl JHEP} {\bf 1202} (2012) 126,
  [\href{http://arxiv.org/abs/1109.3515}{{\sf arXiv:1109.3515}}],
  [\href{http://dx.doi.org/10.1007/JHEP02(2012)126}{{\sf
  doi:10.1007/JHEP02(2012)126}}].

\bibitem{Khosravi:2011zi}
N.~Khosravi, N.~Rahmanpour, H.~R. Sepangi, and S.~Shahidi, {\it {Multi-Metric
  Gravity via Massive Gravity}},  {\sl Phys.Rev.} {\bf D85} (2012) 024049,
  [\href{http://arxiv.org/abs/1111.5346}{{\sf arXiv:1111.5346}}],
  [\href{http://dx.doi.org/10.1103/PhysRevD.85.024049}{{\sf
  doi:10.1103/PhysRevD.85.024049}}].

\bibitem{Akrami:2012vf}
Y.~Akrami, T.~S. Koivisto, and M.~Sandstad, {\it {Accelerated expansion from
  ghost-free bigravity: a statistical analysis with improved generality}},
  {\sl JHEP} {\bf 1303} (2013) 099, [\href{http://arxiv.org/abs/1209.0457}{{\sf
  arXiv:1209.0457}}], [\href{http://dx.doi.org/10.1007/JHEP03(2013)099}{{\sf
  doi:10.1007/JHEP03(2013)099}}].

\bibitem{Akrami:2013ffa}
Y.~Akrami, T.~S. Koivisto, D.~F. Mota, and M.~Sandstad, {\it {Bimetric gravity
  doubly coupled to matter: theory and cosmological implications}},  {\sl JCAP}
  {\bf 1310} (2013) 046, [\href{http://arxiv.org/abs/1306.0004}{{\sf
  arXiv:1306.0004}}],
  [\href{http://dx.doi.org/10.1088/1475-7516/2013/10/046}{{\sf
  doi:10.1088/1475-7516/2013/10/046}}].

\bibitem{Tamanini:2013xia}
N.~Tamanini, E.~N. Saridakis, and T.~S. Koivisto, {\it {The Cosmology of
  Interacting Spin-2 Fields}},  {\sl JCAP} {\bf 1402} (2014) 015,
  [\href{http://arxiv.org/abs/1307.5984}{{\sf arXiv:1307.5984}}],
  [\href{http://dx.doi.org/10.1088/1475-7516/2014/02/015}{{\sf
  doi:10.1088/1475-7516/2014/02/015}}].

\bibitem{Akrami:2014lja}
Y.~Akrami, T.~S. Koivisto, and A.~R. Solomon, {\it {The nature of spacetime in
  bigravity: two metrics or none?}},
  \href{http://arxiv.org/abs/1404.0006}{{\sf arXiv:1404.0006}}.

\bibitem{Yamashita:2014fga}
Y.~Yamashita, A.~De~Felice, and T.~Tanaka, {\it {Appearance of Boulware-Deser
  ghost in bigravity with doubly coupled matter}},
  \href{http://arxiv.org/abs/1408.0487}{{\sf arXiv:1408.0487}}.

\bibitem{Noller:2014sta}
J.~Noller and S.~Melville, {\it {The coupling to matter in Massive, Bi- and
  Multi-Gravity}},  \href{http://arxiv.org/abs/1408.5131}{{\sf
  arXiv:1408.5131}}.

\bibitem{Hassan:2014gta}
S.~Hassan, M.~Kocic, and A.~Schmidt-May, {\it {Absence of ghost in a new
  bimetric-matter coupling}},  \href{http://arxiv.org/abs/1409.1909}{{\sf
  arXiv:1409.1909}}.

\bibitem{Schmidt-May:2014xla}
A.~Schmidt-May, {\it {Mass eigenstates in bimetric theory with ghost-free
  matter coupling}},  \href{http://arxiv.org/abs/1409.3146}{{\sf
  arXiv:1409.3146}}.

\bibitem{Enander:2014xga}
J.~Enander, A.~R. Solomon, Y.~Akrami, and E.~Mortsell, {\it {Cosmic expansion
  histories in doubly-coupled, ghost-free massive bigravity}},
  \href{http://arxiv.org/abs/1409.2860}{{\sf arXiv:1409.2860}}.

\bibitem{deRham:2014fha}
C.~de~Rham, L.~Heisenberg, and R.~H. Ribeiro, {\it {Ghosts and Matter Couplings
  in Massive (bi-and multi-)Gravity}},
  \href{http://arxiv.org/abs/1409.3834}{{\sf arXiv:1409.3834}}.

\bibitem{Weisberg:1981mt}
J.~Weisberg, J.~Taylor, and L.~Fowler, {\it {GRAVITATIONAL WAVES FROM AN
  ORBITING PULSAR}},  {\sl Sci.Am.} {\bf 245} (1981) 66--74,
  [\href{http://dx.doi.org/10.1038/scientificamerican1081-74}{{\sf
  doi:10.1038/scientificamerican1081-74}}].

\bibitem{Gumrukcuoglu:2011ew}
A.~E. Gumrukcuoglu, C.~Lin, and S.~Mukohyama, {\it {Open FRW universes and
  self-acceleration from nonlinear massive gravity}},  {\sl JCAP} {\bf 1111}
  (2011) 030, [\href{http://arxiv.org/abs/1109.3845}{{\sf arXiv:1109.3845}}].

\bibitem{PhysRevD.86.124019}
G.~D'Amico, {\it Cosmology and perturbations in massive gravity},  {\sl Phys.
  Rev. D} {\bf 86} (Dec, 2012) 124019,
  [\href{http://dx.doi.org/10.1103/PhysRevD.86.124019}{{\sf
  doi:10.1103/PhysRevD.86.124019}}].

\end{thebibliography}\endgroup

\end{document}